\documentclass[twocolumn,showpacs,preprintnumbers,amsmath,amssymb]{revtex4}
\usepackage{amsfonts}
\usepackage{amsmath}
\usepackage{graphicx}
\usepackage{dcolumn}
\usepackage{bm}
\usepackage{overpic}
\usepackage{booktabs}

\begin{document}
\title{Enhancing synchronization by directionality in complex networks}
\author{An Zeng$^{1,3}$, Seung-Woo Son$^{2}$, Chi Ho Yeung$^{3}$, Ying Fan$^{1}$ and Zengru Di$^{1}$\footnote{Email: zdi@bnu.edu.cn}}
 \affiliation{$^{1}$Department of Systems Science, School of Management and Center for Complexity
 Research, Beijing Normal University, Beijing 100875, China\\
 $^{2}$Complexity Science Group, Department of Physics and Astronomy, University of Calgary, Calgary, AB, Canada T2N 1N4\\
 $^{3}$Department of Physics, University of Fribourg, Chemin du Musee 3, 1700 Fribourg, Switzerland}

\date{\today}

\begin{abstract}
We proposed a method called residual edge-betweenness gradient (REBG) to enhance synchronizability
of networks by assignment of link direction while
keeping network topology and link weight unchanged. Direction assignment has been shown to improve the synchronizability of undirected networks in general, but we find that in some cases incommunicable components emerge and networks fail to synchronize. We show that the REBG method can effectively avoid the synchronization failure ($R=\lambda_{2}^{r}/\lambda_{N}^{r}=0$) which occurs in the residual degree gradient (RDG) method proposed in Phys. Rev. Lett. 103, 228702 (2009). Further experiments show that REBG method enhance synchronizability in networks with community structure as compared with the RDG method.


\end{abstract}


\maketitle

Synchronization is an important phenomenon in various fields including biology, physics, engineering, and even sociology~\cite{PR93}. In
particular, synchronization in complex networks has been
intensively studied in the past
decade~\cite{PRL014101,PRE067105,PRE015101,PRL034101,PRL0341012,PNAS10342,PRL218701,PRL138701,PRE016116}.
One important objective in these studies is to
enhance the
synchronizability~\cite{PRE016116,PRL034101,PRL218701,PRL138701},
i.e., the ability to coordinate oscillators in synchronization.
Many related methods based on node properties and link weight have
been proposed. For example, some researchers took into account the
degree centrality~\cite{PRE016116,PRL034101}, the
betweenness~\cite{PRL218701} and the node
age~\cite{PRL138701,PRE057103}, and assigned weight on links to
enhance synchronizability. Nishikawa and Motter proposed to assign
zero weight to particular links which lead to an oriented tree with
normalized input strength and no directed
loop~\cite{PRE065106,PhysicaD77}. Moreover, they proved that tree
is the optimal structure on which synchronizability is highest.
Recently, the shortest oriented spanning tree is shown to be the optimal structure
for both synchronizability and the convergence
time~\cite{EPL48002,NJP043030}. However, these methods are all
based on assigning weights to links, where the influences of link
direction on synchronization have not been intensively
considered~\cite{PR93}.

How to improve synchronization in directed network is still an
unsolved problem. Though previous works suggest that hierarchical
structure and the absence of feed-back loop can enhance
synchronizability, the underlying mechanism is not clear. On the
other hand, the directionality plays a significant role in the
dynamic of
networks~\cite{PRL118701,PRE046121,PRE016106,PRE026114}. With the
understanding of relations between link direction and
synchronization, a lot of applications can be made~\cite{PR93}.
For example, simply regulating the direction of the phase signal
of alternating current can facilitate phase match in power grids
without additional construction cost in the topology. With this in
mind, the authors in Ref.~\cite{PRL103} proposed the residual
degree gradient (RDG) method to enhance directed networks'
synchronizability by assigning only the direction of links without
changing the entire topology and link weights. They also claimed
that RDG method can enhance the network synchronization, contrary
to the randomly assigned directional method.

However, instead of enhancing synchronizability, we find that the
RDG method results in incommunicable components in some
particular cases, which lead to graphs incapable of complete synchronization.
Incommunicable components of a network correspond to the
components on which information cannot be transmitted from one to
the other in either direction. Under these circumstance, the
networks can never reach complete synchronized state. In this
paper, we propose the so-called residual edge-betweenness gradient
(REBG) method to resolve the problem of incommunicable components
in the RDG method. By evaluating the betweenness on all edges, we
devise an algorithm to assign link direction. The effectiveness of
the algorithm lies in the use of edge betweenness, which embeds
global information, as compared to the node degree which reflect
only local information. We find that REBG is effective in
enhancing synchronizability without leading to incommunicable
components in most network topologies.

To begin our analysis, we make use of
the Master Stability Function which allows us to use the
eigenratio $R=\lambda_{\rm min}/\lambda_{\rm max}$ of the Laplacian matrix
to represent the synchronizability of a network, with
$\lambda_{\rm min}$ and $\lambda_{\rm max}$ denote respectively the
smallest nonzero eigenvalue and the largest
eigenvalue~\cite{PRL054101,PRL2109,PRE5080}. Specifically, $0 \leq
R \leq 1$, the larger the value of $R$, the stronger the
synchronizability of networks. According to
Ref.~\cite{PRL103,PRL138701}, we use the real part of eigenvalues
from the Laplacian matrix to investigate the propensity for
synchronization in directed networks. However, instead of setting
$R=\lambda_{\rm min}^{r}/\lambda_{\rm max}^{r}$ (superscript $r$ denotes
the real part of complex number), we set
$R=\lambda_{2}^{r}/\lambda_{N}^{r}$, with $\lambda_{2}^{r}$ to be
the second smallest eigenvalue. It is because $\lambda_{2}^{r}=0$
in cases when incommunicable components emerge, which is not
reflected by $\lambda_{\rm min}^{r}$. In the subsequent analysis, we
use $R$ to characterize synchronizability.

As the known fact, when there is isolated node or community in an
undirected network, the network can never reach complete
synchronization since there is no information flow between the
isolated components. Under this circumstances, the
synchronizability index $R=\lambda_{2}/\lambda_{N}=0$. We call
this phenomenon ($R=0$) \emph{synchronization failure}. Generally
speaking, synchronization failure is more likely in directed
networks as isolated components are not necessary. We first denote
the \emph{cut-vertex} as the only node through which two or more
components communicate with each other. In directed networks, the
synchronization failure happens whenever cut-vertex have only
incoming links. In this case, there is absolutely no communication
between those components and the cut-vertex becomes an information
sink. Hence, synchronization cannot be achieved among the
incommunicable components.

In order to examine synchronization failure, we first describe
briefly the RDG method~\cite{PRL103}. The RDG method is proposed
to enhance the synchronizability of undirected networks by simply
assigning the link direction. They refer links without yet
assignment of direction as residual edges, and the number of
connected residual edges as the residual degree of a node. In each
step, they select the node with the smallest residual degree and
set a maximum of $\lceil\langle k\rangle/2\rceil$ of its residual
edges pointing to it\cite{explanation}, where $\langle k \rangle$ is the average
degree in the original network. Once a node has been selected, it
will not be chosen again. Nodes have not yet been selected are
called residual nodes. In addition, there is a {\em directionality}
$\alpha$ to control the final fraction of links with direction
assignment. When $\alpha=0$, all links remain undirected. When
$\alpha=1$, all links are assigned direction. The RDG assignments
are finished when there is no residual node left in the network.

The RDG method may lead to directed graph with synchronization
failure, i.e., cut-vertex with only incoming links. A simple
example is given in Fig. 1 (a). According to the rule of RDG, node
1 ($k=2$) will be selected first and the two remaining communities
will be left incommunicable. As we have discussed, this RDG
network can never reach complete synchronized state.

\begin{figure}
  \center
  \includegraphics[width=5.5cm]{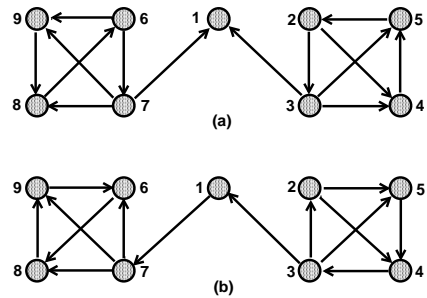}
  \caption{(a) A simple example of RDG network with resultant $R=0$ and original $R=0.039$ (b) The REBG network from the same
  original network with $\theta=1$, and resultant $R=0.4$.}
\end{figure}

\begin{figure}
  \center
  \includegraphics[width=4.27cm]{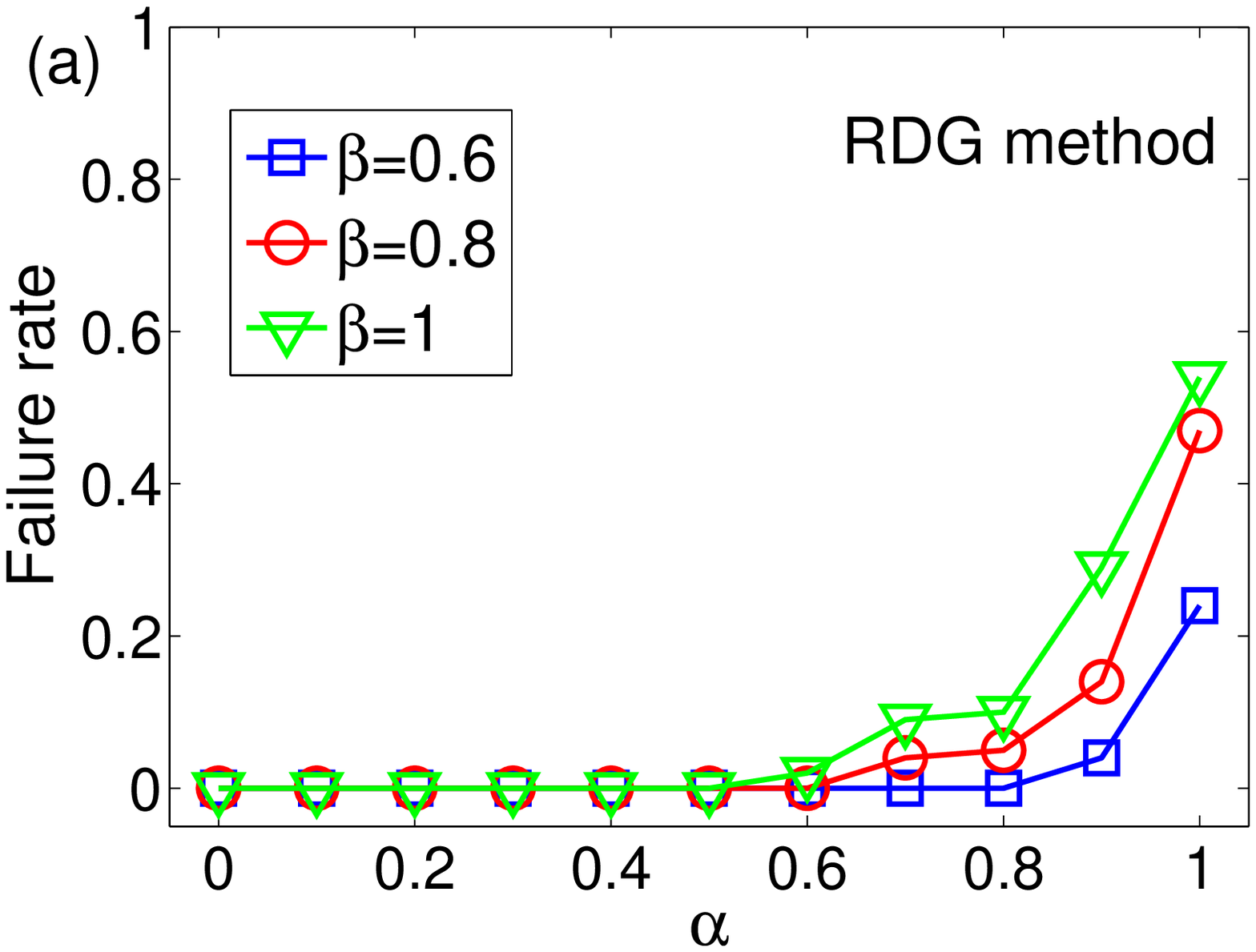}
  \includegraphics[width=4.27cm]{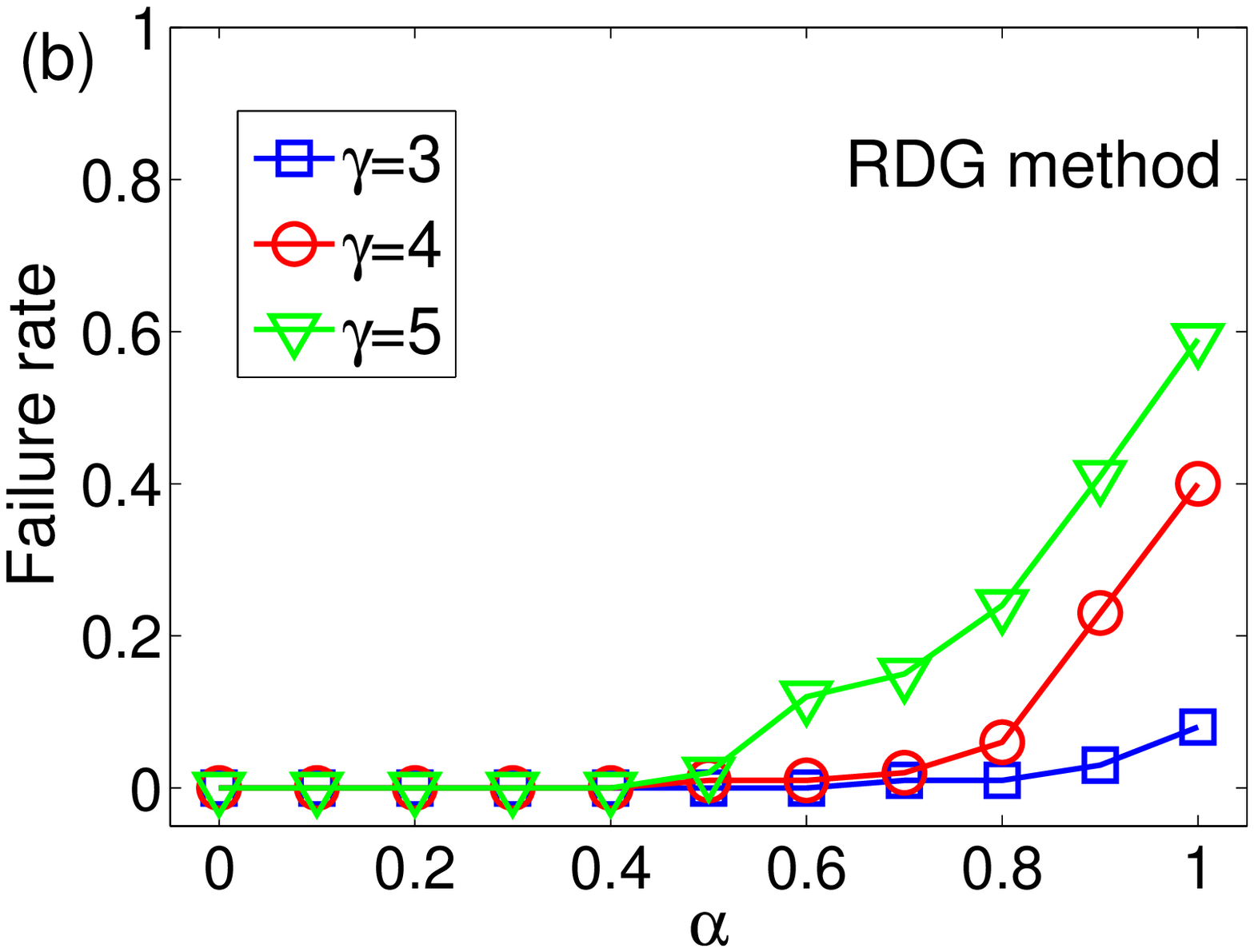}\\
  \includegraphics[width=4.27cm]{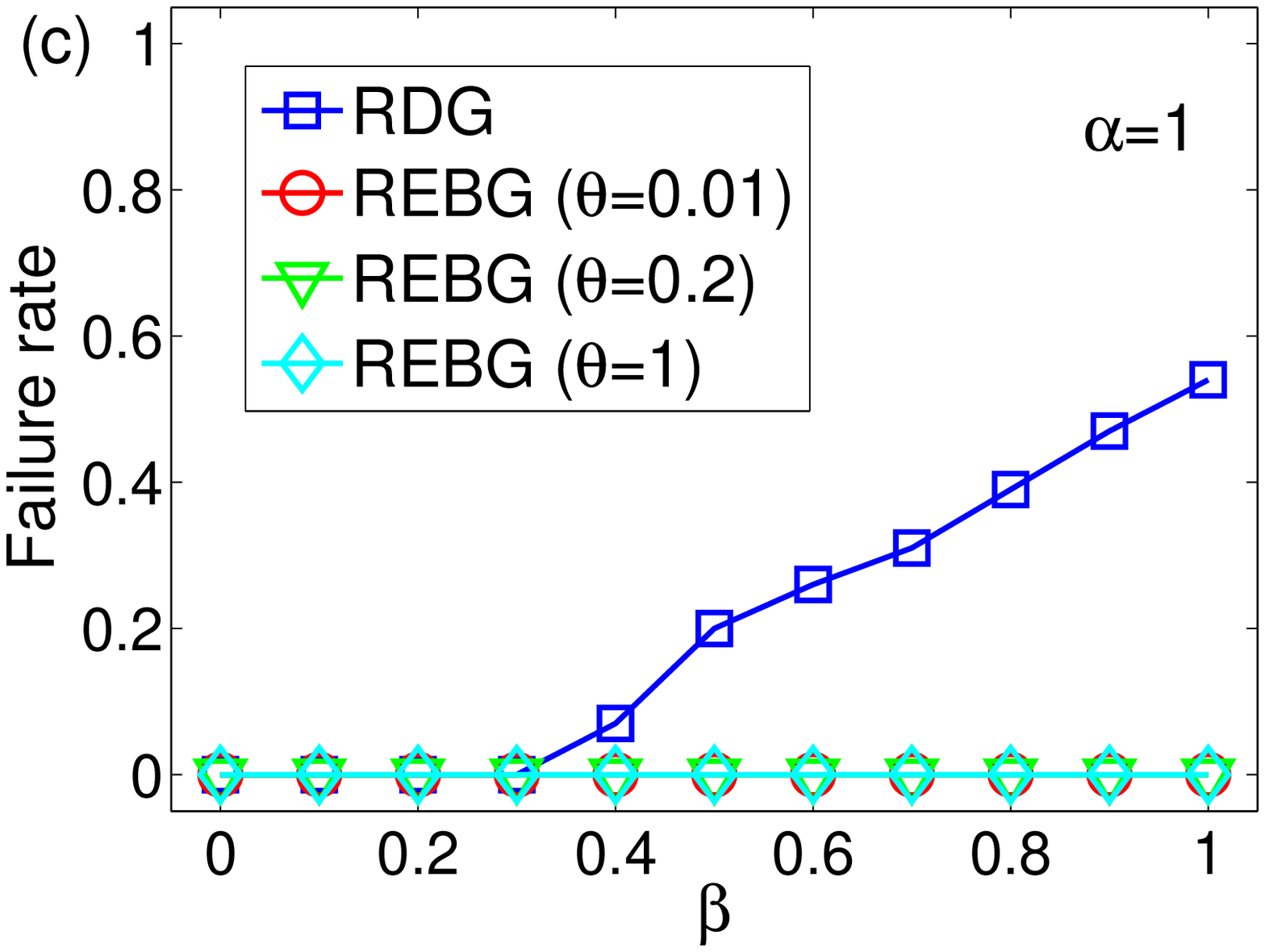}
  \includegraphics[width=4.27cm]{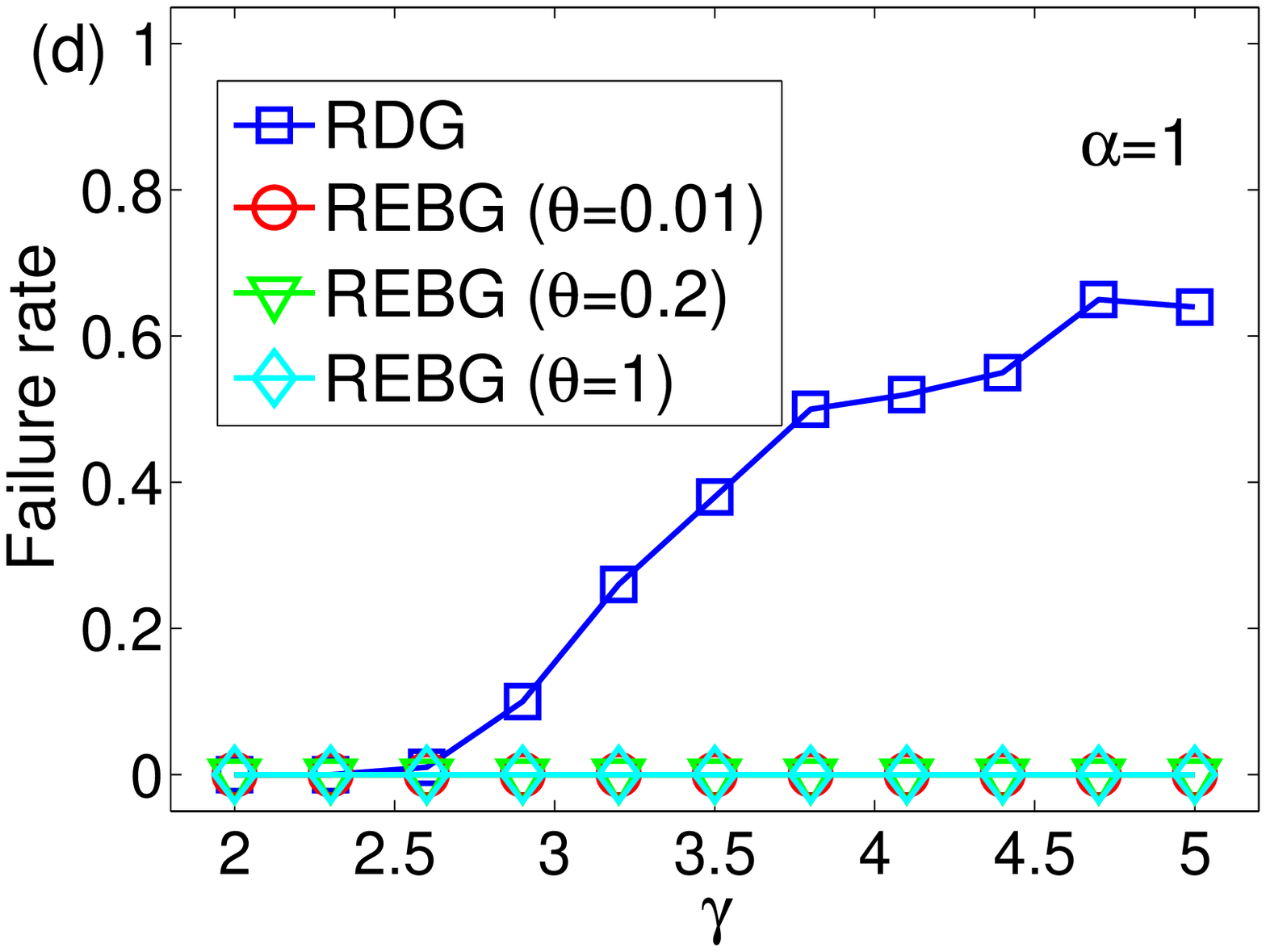}
  \vspace{-0.5cm}
  \caption{(Color online) The synchronization failure rate as a function of $\alpha$ by the RDG method in (a) random exponential networks ($P(k)\thicksim e^{-\beta k}$, $N=500$, $k_{\rm min}=2$, $\beta=1$) and (b) random scale-free networks ($P(k)\thicksim k^{-\gamma}$,
  $N=500$, $k_{\rm min}=2$, $\gamma=5$). Given $\alpha=1$, the failure rate as a function of $\beta$ in (c) random exponential network and the failure rate as a function of $\gamma$ in (d) random scale-free networks when RDG and REBG ($\theta=0.01$, $\theta=0.2$ and $\theta=1$) methods are used.}
\end{figure}

According to~\cite{PRL103}, RDG was supposed to provide a
practical way for the synchronization improvement in power grid
networks, which have exponential degree distribution~\cite{PRE69}.
So we first study the RDG method in \emph{random exponential
networks}~\cite{PRE64}. Random exponential networks in our context
refer to random networks with exponential degree distribution.
We also tested the RDG method in \emph{random scale-free
networks} since power-law degree distributions are widely observed in
empirical data~\cite{PRE64}. The degree distribution are
respectively given by $P(k)\thicksim e^{-\beta k}$ and
$P(k)\thicksim k^{-\gamma}$. For each $\beta$ and $\gamma$, we
tested the RDG method on 100 network realizations. To examine
synchronization failure, we make sure that there is no isolated
nodes or communities in the original networks which masks the
effect of incommunicable components. We find that many resultant
RDG networks are with $R=0$.
The failure rate,
i.e. the fraction of realization incapable of complete synchronization,
is reported in Fig. 2.
As shown in Fig. 2(a) and (b), when
$\alpha$ increase (i.e., more links are assigned direction),
failure rate increases for both exponential and power-law
networks. These results imply that synchronization failure is
common in both power-grid-like and scale-free networks, given RDG
is used as the method for direction assignment. We further tested
the case of $\alpha=1$, which leads to the largest synchronization
improvement as reported in~\cite{PRL103}. By varying $\beta$ and
$\gamma$, we find that failure is high when $\beta$ and $\gamma$
are large.
We also remark that failure rate decreases with $k_{\rm min}$
but increases with $N$.
All these suggest that synchronization failure is not a
rare phenomenon and therefore a new method of direction assignment
is required to prevent failure.

We thus introduce the residual edge-betweenness gradient (REBG) method
to solve the problem of synchronization failure.
Instead of the node degree, we take the edge-betweenness into account.
First of all, we define $s_{i}$ for node $i$ as
\begin{equation}
s_{i}=\sum_{j=1}^{N}a_{ij}l_{ij}^{\theta},
\end{equation}
where $a_{ij}=1$ when there exists an \emph{undirected} link between $i$ and $j$ and otherwise 0. 
$l_{ij}$ is the betweenness of the link between $i$ and $j$
evaluated on the original undirected networks,
subjected to a power $\theta$ with $0 \leq\theta\leq 1$.
To assign link direction, we select
the node with the smallest residual $s_{i}$ in each step and
assign an incoming direction for a maximum of $\lceil\langle
k\rangle/2\rceil$ of its residual links. As more directed links are assigned,
the residual $s_{i}$ has to be updated at every step. If there are
multiple nodes of the smallest $s_{i}$, we choose the node with
the smallest initial $s_{i}$ first. The REBG method stops when
there is no residual node left in the network. We remark that when
the parameter $\theta=0$, $s_{i}=k_{i}$ and the REBG reduces to
the RDG method. When $\theta>0$, the REBG method contains the global information delivered from the edge betweenness.

\begin{figure}
  \center
  \includegraphics[width=4.27cm]{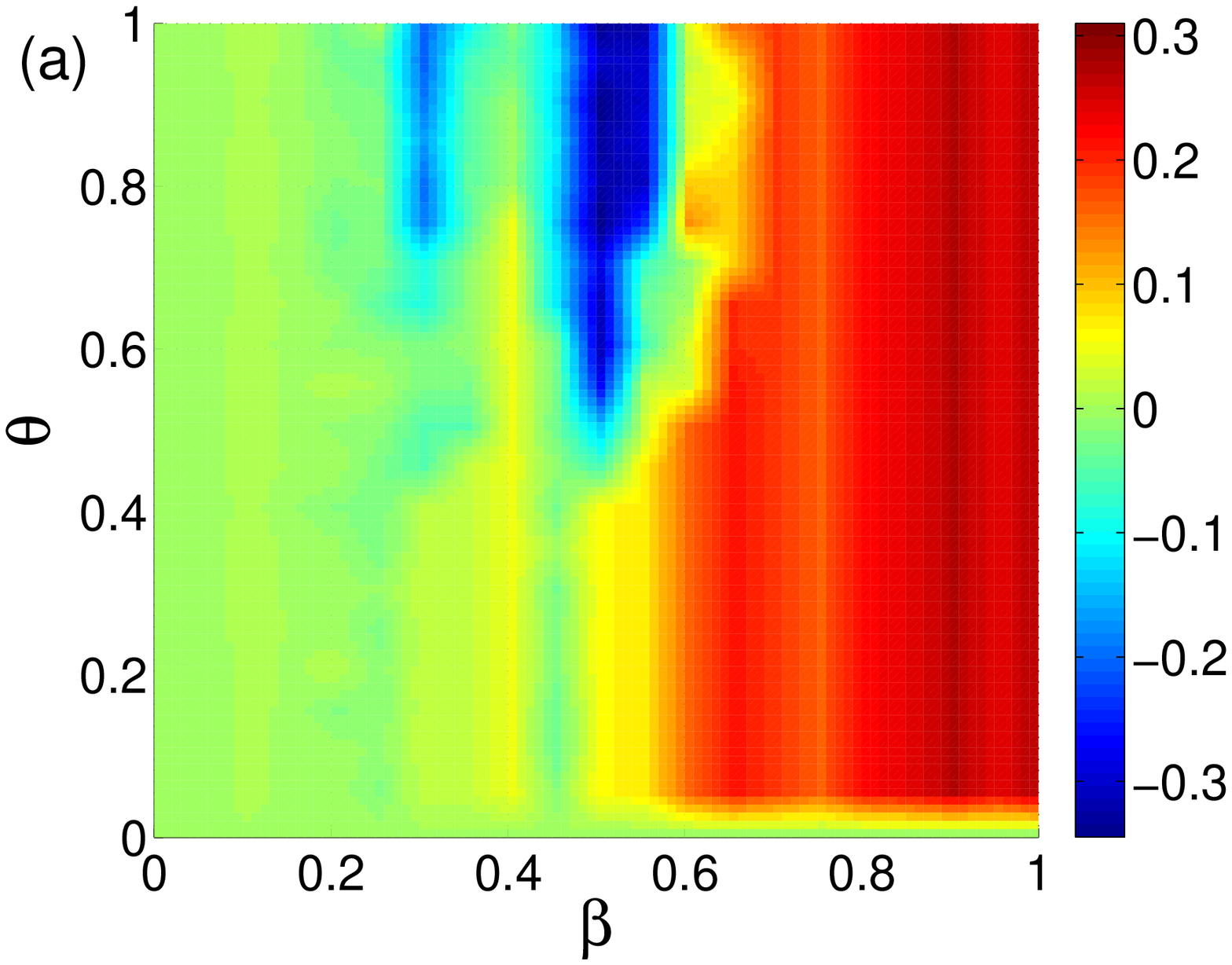}
  \includegraphics[width=4.27cm]{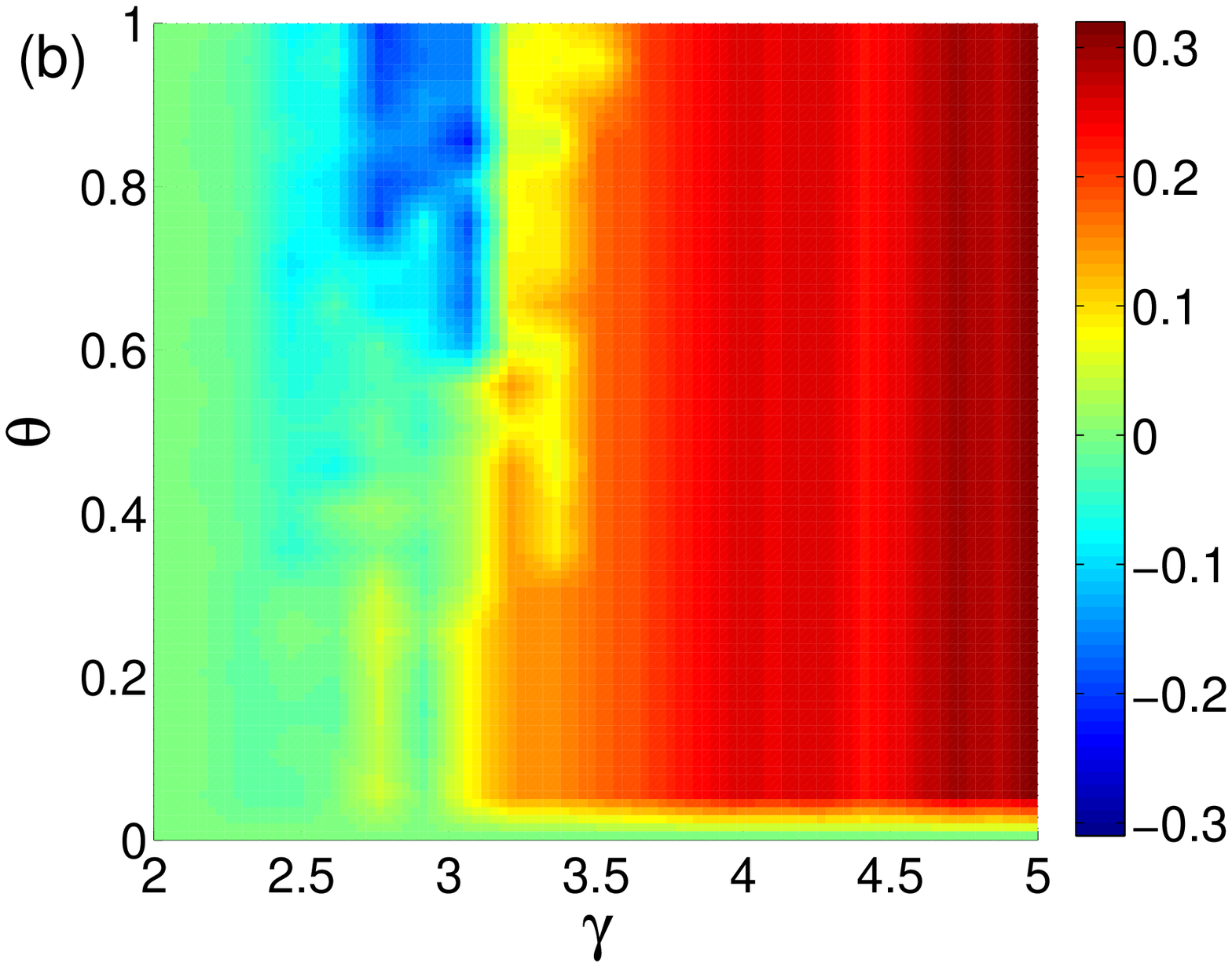}
  \vspace{-0.5cm}
  \caption{(Color online) The difference $D=R_{\rm REBG}-R_{\rm RDG}$ in
  (a) random exponential networks
  and (b) random scale-free networks with $N=500$ and $k_{\rm min}=2$. The results are obtained by averaging 10 independent realizations. }
\end{figure}

\begin{figure}
  \center
  \includegraphics[width=4.27cm]{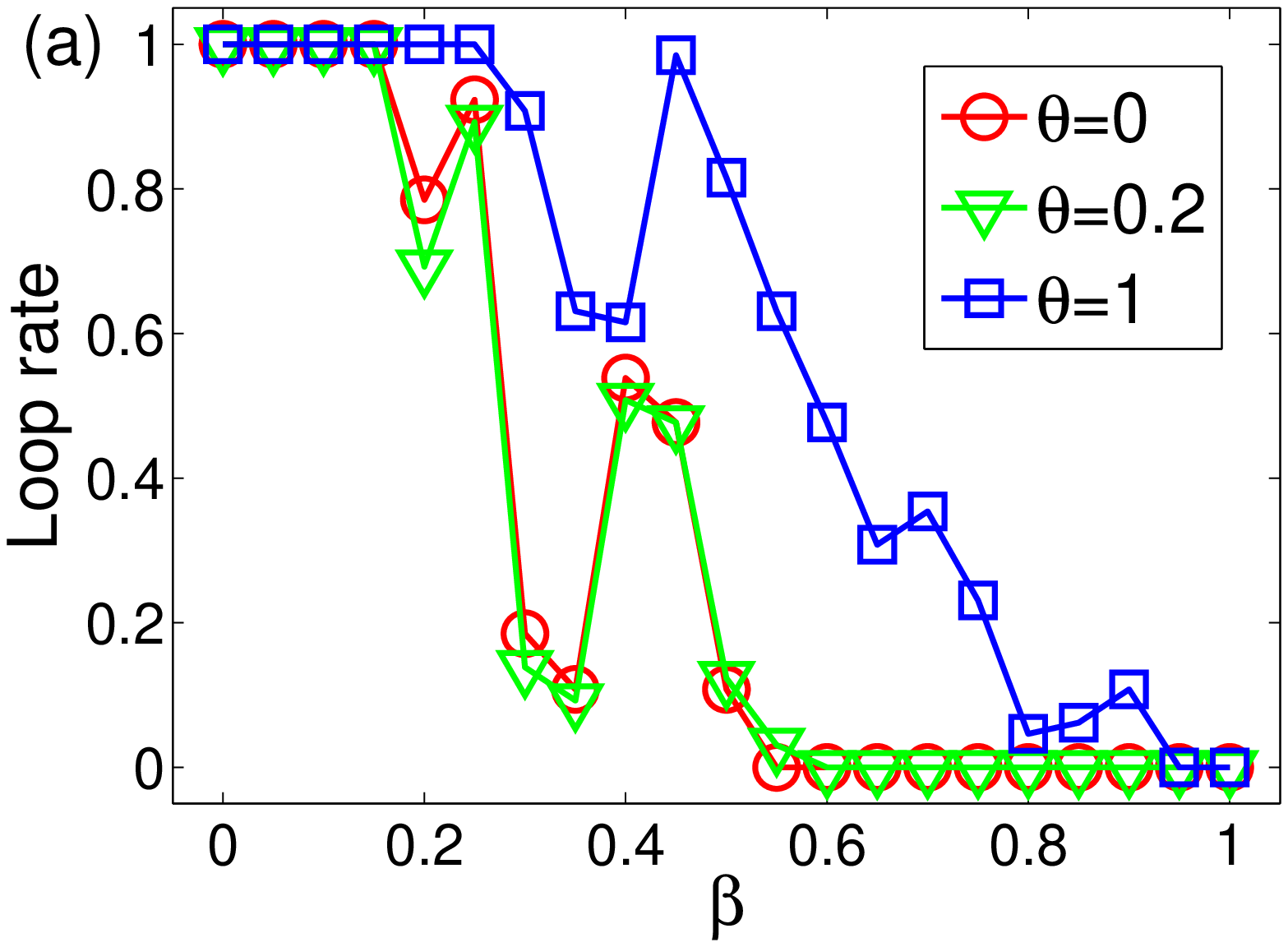}
  \includegraphics[width=4.27cm]{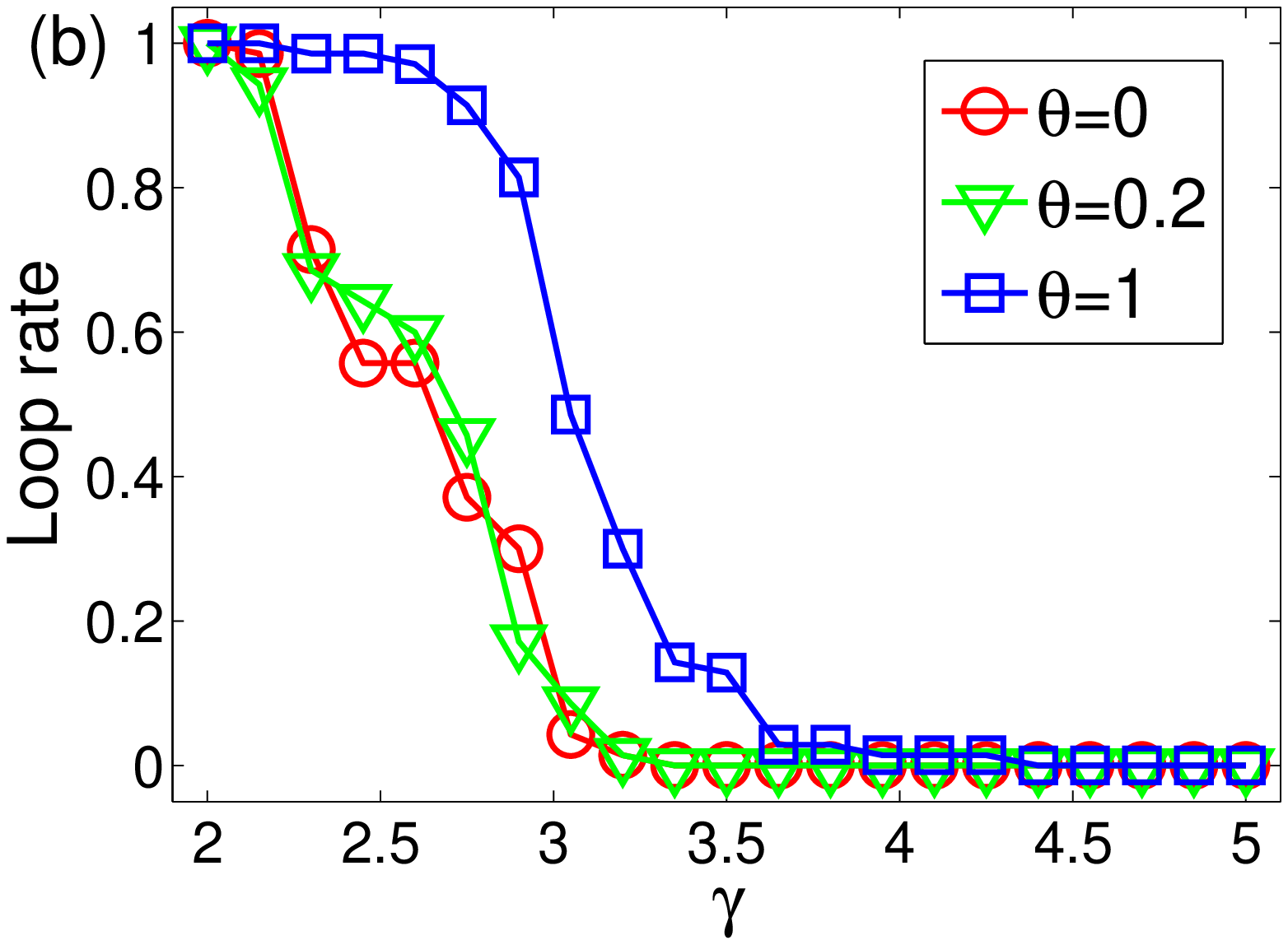}
  \vspace{-0.5cm}
  \caption{(Color online) The fraction of realizations with directed loop as obtained by the REBG and RDG method in (a) random exponential networks and (b) random scale-free networks. The results are obtained by average 100 realizations. The abrupt jumps in random exponential networks come from the discontinuity in the assignment constraint $\lceil\langle k\rangle/2\rceil$.}
\end{figure}

An shown in Fig. 1 (b), the REBG method can effectively avoid the
problem of synchronization failure. In this simple network,
whenever node $1$ is chosen before node $3$ and $7$,
synchronization failure occurs. We note that node $1$ is the
cut-vertex connecting the two communities, its edges are of high
betweenness. To see how failure is avoided, we evaluate the
initial $s_{i}$ before any direction assignment, as given by
$s_{1}=20^{\theta}+20^{\theta}$,
$s_{x}=1^{\theta}+1^{\theta}+6^{\theta}$ for $x=2,4,5,6,8,9$ and
$s_{y}=6^{\theta}+6^{\theta}+6^{\theta}+20^{\theta}$ for $y=3,7$.
By tracing all the possibilities of subsequent direction
assignment, we find that synchronization failure does not occur
provided that node $1$ is not selected at the first assignment. In
this case, $s_{1}>s_{x}$ which implies $\theta\gtrsim 0.18$. We
denote this value as $\theta_{c}$ which marks the value of
$\theta$ from which failure ceases. We
remark that the value of $\theta_{c}$ is different for different
topology.

We then examine the failure rate in the random exponential
networks and the random scale-free networks. As shown in Fig. 2(c) and (d), failure rate vanishes for both networks when
$\theta=0.01$, $0.2$ and $1$. These results suggest
$\theta_{c}\approx 0$, i.e., failure ceases when edge betweenness
is considered in direction assignment. It implies that $\theta$ in
Eq. (1) should be positive to make complete
synchronized state possible.

Finally, we compare the synchronizability index $R$ between REBG
and RDG methods. Both random exponential networks and random
scale-free networks are examined. For better illustration, we
report the difference $D=R_{\rm REBG}-R_{\rm RDG}$ as a function of
$\theta$ and $\beta$ in Fig. 3(a), and $\theta$ and $\gamma$ in Fig.
3(b). The positive $D$ shows that the REBG method results in higher
synchronizability as compared to the RDG method. This
enhancement is mainly resulted from the
prevention of the synchronization failure. One may question the
validity of indicating synchronizability by $R$ with only real
part of eigenvalues. We show in Fig. 4 that for the regime when
$\beta$ and $\gamma$ is large, the resultant networks are free of
directed loops, which support the validity of positive $D$ (as
obtained by $R$) in this regime. These results also explain the
negative $D$ observed with intermediate value of $\beta$ and
$\gamma$. From the lines of $\theta=0$ and $\theta=1$, we see that
the number of networks with directed loops is higher when
$\theta=1$, which hinders synchronization and lead to unfavorable
results from REBG. We further show that the REBG method with
$\theta=0.2$ are similar to the RDG in terms of the fraction of
loopy realizations, suggesting $\theta=0.2$ is an effective value
for direction assignment, and at the same time avoids synchronization failure.
Hence, one can use the REBG method with small
$\theta$ to enhance synchronizability effectively.

\begin{figure}
  \center
  \vspace{-0.5cm}
  \includegraphics[width=9cm]{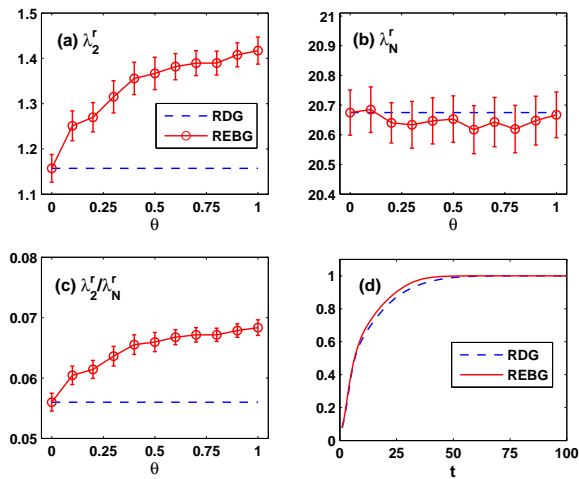}
  \vspace{-1cm}
  \caption{(Color online) The synchronizability of the REBG method and the RDG method in GN-benchmark($k_{\rm outer}=1$) as measured by (a) $\lambda_{2}^{r}$, (b) $\lambda_{N}^{r}$ and (c) $R=\lambda_{2}^{r}/\lambda_{N}^{r}$ as a function of $\theta$. (d) The order parameter $r(t)$ of Kuramoto model ($\sigma=10$) on the RDG networks and REBG networks($\theta=1$). The results are obtained by averaging 100 independent realizations. }
\end{figure}

We further compare the RDG and the REBG methods in graphs where failure
does not occur.
Here, we study the two methods in Girvan Newman benchmark
(GN-benchmark) network which consists of $128$ nodes and is
divided into 4 communities~\cite{PRE026113}. In GN-benchmark,
$k_{\rm inter}+k_{\rm outer}=16$, where $k_{\rm inter}$ is the average node
degree in each community and $k_{\rm outer}$ is the average node
degree among different communities. We show in Fig. 5 the results
of $k_{\rm outer}=1$ from which the GN-benchmark networks are highly
clustered. We can see from Fig. 5(a) that the synchronization
enhancement comes mainly from $\lambda_{2}^{r}$. Moreover, we
tested the Kuramoto model on the resultant RDG networks and REBG
networks. The oscillator on node $i$ of the networks is described
by $\dot{\theta_{i}}=\omega_{i}+\sigma\sum_{j}^{N}
A_{ij}\sin(\theta_{j}-\theta_{i})$ in which $A$ is the adjacency
matrix and $A_{ij}$ denote the information flow from $j$ to $i$,
and the collective phase synchronization can be investigated by
the order parameter defined as
$r(t)=\langle|(\sum_{j=1}^{N}e^{i\theta_{j}(t)}/N)|\rangle$. From
Fig. 5(d), it is obvious that $r(t)$ of the REBG networks
converges faster than that of the RDG networks. These results show
that the REBG method lead to greater improvement in
synchronizability than the RDG method in highly clustered
networks, despite the absence of failure in RDG networks.

In summary, we introduced the
residual edge betweenness gradient (REBG) method for direction
assignment, which overcomes the problem of emergence of
incommunicable components in the residual degree gradient (RDG)
method. The effectiveness of our method lies in the use of edge
betweenness, which reflects global network information when
compared to the node degree in RDG. Further tests of REBG and RDG
in highly clustered networks show that REBG can lead to greater
synchronizability improvement in networks despite the absence of
failure problem. For daily applications, such
incommunicable components brings huge loss to the electrical
systems when power grids are unable to reach complete
synchronization. Hence, the REBG method is effective in improving
synchronizability which may lead to wide applications.

This work was supported by NSFC under Grant No. 60974084 and No.
70771011. CHY is partially supported by the QLectives projects (EU
FET-open Grants 213360 and 231200).


\begin{thebibliography}{99}
\bibitem{explanation} Different from~\cite{PRL103}, please note that the
edge direction here is the direction of information flow.
\end{thebibliography}
\end{document}